\documentclass[preprint,showpacs,aps]{revtex4}
\usepackage{graphicx}% Include figure files
\begin{document}

\title{Edge dislocations in crystal structures
considered as traveling waves of discrete models}
\author{A. Carpio$^1$ and L. L. Bonilla$^2$ }
\affiliation{$^1$Departamento de Matem\'{a}tica
Aplicada, Universidad Complutense de Madrid,\\
28040 Madrid, Spain\\
$^2$Departamento de Matem\'{a}ticas, Universidad Carlos III
de Madrid,\\ Avenida de la Universidad 30, 28911
Legan{\'e}s, Spain}
\date{ \today  }

\begin{abstract}
The static stress needed to depin a 2D edge dislocation, the
lower dynamic stress needed to keep it moving, its velocity
and displacement vector profile are calculated from first
principles. We use a simplified discrete model whose far
field distortion tensor decays algebraically with distance as
in the usual elasticity. Dislocation depinning in the strongly
overdamped case (including the effect of fluctuations) is
analytically described. $N$ parallel edge dislocations
whose average inter-dislocation distance divided by the Burgers
vector of a single dislocation is $L\gg 1$ can depin a given one
if $N=O(L)$. Then a limiting dislocation density can be
defined and calculated in simple cases. 
\end{abstract}
\pacs{61.72.Bb, 5.45.-a, 45.05.+x, 82.40.Bj}
%\begin{multicols}{2}
\maketitle
%\narrowtext

In many fields, genuinely microscopic phenomena affect
macroscopic behavior in a way that is difficult to quantify
precisely. Typical cases are the motions of dislocations
\cite{nab67,hir82}, cracks \cite{fre90}, vortices in Josephson
arrays \cite{zan95} or other defects subject to pinning due to
the underlying crystal microstructure. Emerging behavior due to
motion and interaction of defects might explain common but
poorly understood phenomena such as friction \cite{ger01}.
Macroscopic theories consider the continuum mechanics of these
solids subject to forces due to the defects and additional
equations for the densities of defects and properties of their
motion \cite{ll7}. The latter are usually postulated by
phenomenological considerations. An important problem is to
derive a consistent macroscopic description taking into account
the microstructure. 

Here we tackle a simplified problem containing all the
ingredients of the previous description: the pinning and motion
of edge dislocations. Firstly, we study a two-dimensional (2D)
discrete model \cite{kkl} describing the damped displacement of
atoms subject to the field generated by a 2D edge dislocation
and a constant applied shear stress of strength $F$. If $|F|<
F_{cs}$ ($F_{cs}$ is related to the static Peierls stress),
the stable displacement field is stationary, whereas the 
dislocation core and its surrounding displacement field move if
$|F|>F_{cs}$. A crucial observation is that there exists a
stable uniformly moving dislocation with both core and far field
advancing at the same constant velocity. This suggests that
{\em a moving edge dislocation is a traveling wave of the
discrete model}. 

Our self-consistent calculation \cite{self} based on this
picture predicts the following magnitudes: (i) the critical
static stress needed to depin a static dislocation, (ii) the
dynamic stress $F_{cd}<F_{cs}$ below which a moving dislocation
stops (in the strongly overdamped case, $F_{cd}=F_{cs}$), and
(iii) the dislocation velocity as a function of applied stress.
The latter information has to be taken from experiments in the
standard treatment \cite{hul01}. Then a macroscopic quantity,
the dislocation velocity, is obtained from analysis of a
microscopic model. 

Secondly, we consider a distribution of many parallel edge
dislocations separated by macroscopic distances comprising many
lattice periods. A dislocation cannot move under the influence
of other dislocations far away unless the latter have finite
density (there are $N$ such dislocations and the average
distance between them is $L$ with $N=O(L)$ as $L\to\infty$).
Under the influence of such a distribution, one dislocation may
be pinned or move depending on the dislocation density. The
latter is calculated at the critical stress in a simple
configuration.

{\em A simplified discrete model of edge dislocations.} Consider
an infinite three dimensional cubic lattice with symmetry axes
$x,y,z$. We insert an extra half plane of atoms parallel to the
plane $yz$. The border of this extra half plane is a line
(parallel to the $z$ axis), which is called an edge dislocation
in the crystal. The Burgers vector of the dislocation points in
the $x$ direction and the plane $xz$ is the glide plane of the
dislocation (see \cite{ll7}). If we apply an external stress
$\sigma$, the dislocation moves on its glide plane and in the
direction of its Burgers vector in response to just one
component of $\sigma$: the stress $\tau$ resolved on the glide
plane in the glide direction \cite{hir82,ll7}. All the sections
of the lattice by planes parallel to $xy$ look alike. Thus, we
reduce the problem to a 2D lattice with an extra half line of
atoms, see Fig.~22 of Ref. \onlinecite{ll7}. Assuming that glide
is only possible in the $x$-direction, the dynamics of an edge
dislocation in a 2D lattice can be described by \cite{kkl}:
\begin{eqnarray}
m\, {d^2u_{i,j}\over dt^2} &+& {du_{i,j}\over dt} = u_{i+1,j}-
2u_{i,j} + u_{i-1,j} \nonumber\\
&+& A\, [\sin(u_{i,j+1}-u_{i,j})+\sin(u_{i,j-1}-u_{i,j})] .
\label{kkl}
\end{eqnarray}
The lattice is a collection of chains in the $x$ direction with
elastic interaction between nearest neighbors within the same
chain and sinusoidal interaction between chains. $u_{i,j}/(2
\pi)$ is the dimensionless displacement of atom $(i,j)$ in the
$x$ direction measured in units of the Burgers vector length
$b$. $A>0$ measures the relative strengths of the nonlinear
forces exerted by atoms on different planes $y=k$ (constant) 
and the linear forces exerted within any plane $y=k$. The
dimensionless parameter $A$ also determines the width of the
dislocation core ($\sim 1/\sqrt{A}$). Lastly, the time unit is
the ratio between the friction coefficient and the spring
constant in the $x$ direction. Then $m$ is the dimensionless
ratio between the atomic mass times the spring coefficient and 
the square of the friction coefficient. In dislocation
dynamics, an important case is that of overdamped dynamics,
$m=0$ \cite{gro00}. Eq.\ (\ref{kkl}) can be generalized to a
vector model having a displacement vector $(u_{ij}, v_{ij})$
and a continuum limit yielding the 2D Navier equations with
cubic symmetry \cite{car03}. Such model has among its solutions
edge dislocations with Burgers vectors in the $x$ or $y$
directions gliding in the direction thereof (which does not
have to be assumed as in the present simple model). This model
can also be solved using our methods at the expense of
technical complications and high computational cost.

In this geometry, the far field of a static 2D edge dislocation
is approximately given by the corresponding continuum elastic
displacement $u_{ij}= u(x,y)$ with $x=\epsilon i$, $y=\epsilon
j$ (where $\epsilon=b/L\ll 1$, $i$, $j$ are large and $L$ is the
appropriate mesoscopic length). Then the stationary solutions of
Eq.\ (\ref{kkl}) satisfy the equations of anisotropic linear
elasticity, $u_{xx} + A  u_{yy}=u_{xx} + u_{YY}=0$, ($Y=y/
\sqrt{A}$) far away from singularities and jumps \cite{kkl}.
The solution corresponding to the edge dislocation is the polar
angle $\theta(x,Y) \in [0,2\pi)$, measured from the positive $x$
axis. Continuum approximations break down near the dislocation
core, which should be described by the discrete model
\cite{car97}. The advantadge of Eq.\ (\ref{kkl}) compared to
other 2D generalizations \cite{lom86} of the Frenkel-Kontorova
model is that it yields the correct decay for strains and
stresses: $r^{-1}$ as $r^2= x^2+Y^2 \to \infty$, instead of
exponential decay.

\begin{figure}
\begin{center}
\includegraphics[width=8cm]{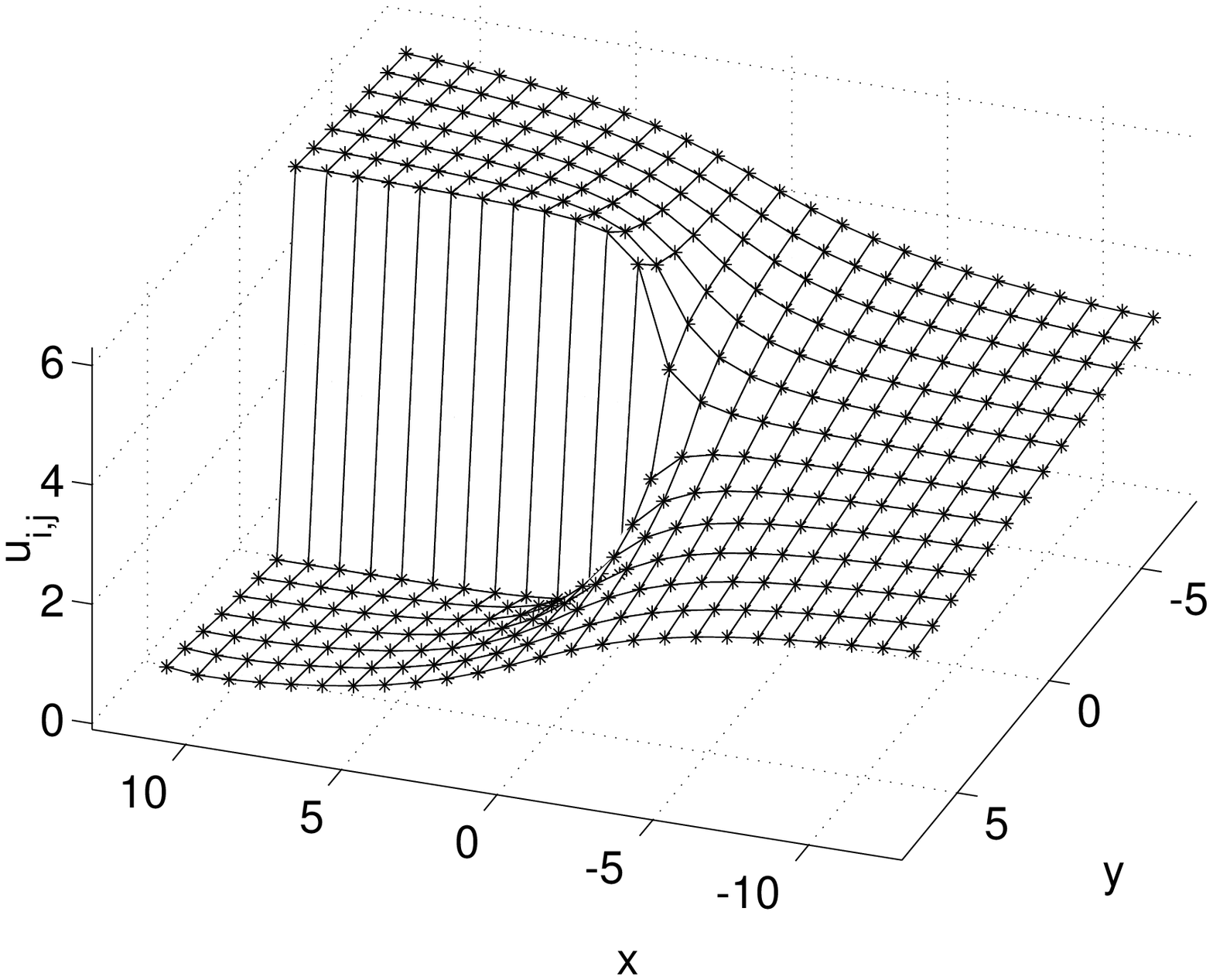}
\caption{Displacement field profile for the stationary edge
dislocation with $A=1$ and $N=50$.}
\label{fig2}
\end{center}
\end{figure}

{\em Overdamped dynamics and static Peierls stress.} We shall
now study the structure of a static edge dislocation of Eq.\
(\ref{kkl}), the critical stress needed to set it in motion and
its subsequent speed. We solve numerically Eq.\ (\ref{kkl})
with $m=0$ on a large lattice $|i|,|j|\leq N$ using Neumann
boundary conditions (NBC) corresponding to applying a shear
stress of strength $F$ in the $x$ direction. The (far field)
continuum elastic displacement for a static 2D edge dislocation
subject to such a shear stress is $\theta(x,y/\sqrt{A})+ Fy$.
Then the NBC are $u_{\pm (N+1),j} - u_{\pm N,j} = \pm [\theta
^A_{\pm (N+1),j} - \theta^A_{\pm N,j}]$ and $u_{i,\pm (N+1)} -
u_{i,\pm N} = \pm [\theta^A_{i,\pm (N +1)} - \theta^A _{i,\pm
N}] \pm F$, where $\theta^A_{i,j}= \theta(i,j/\sqrt{A})$ with
$\theta(0,0)=\pi/2$. If $F=0$ and the initial condition is the
elastic far field $\theta^A_{i,j}$, the system relaxes to a
stationary configuration $u_{i,j}$. The dislocation is expected
to remain stationary for $F\neq 0$ unless $|F|$ is larger than
a critical value $F_{cs}(A)$, related to the so called static
Peierls stress \cite{nab67,hir82}. Nonlinear stability of the
stationary edge dislocation for $|F|<F_{cs}$ was proven in Ref.
\onlinecite{car02}. To test this picture, we solve numerically
Eq.\ (\ref{kkl}) in  a large lattice, using NBC and the static
dislocation obtained for $F=0$ as initial data. For large times
and $|F|$ small, the system relaxes to a steady configuration
$u_{i,j}$ which provides the structure of the core, see Fig.
\ref{fig2}. When $|F|$ is large enough, the dislocation is
observed to glide in the $x$ direction: to the right if $F>0$,
and to the left if $F<0$. 

\begin{figure}
\begin{center}
\includegraphics[width=8cm]{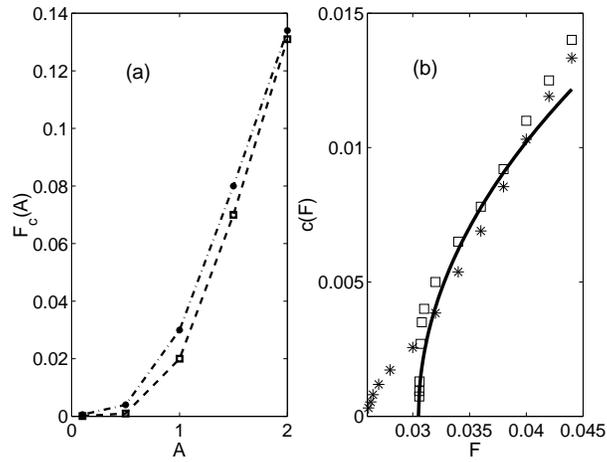}
\caption{(a) Static (squares, $m=0$) and dynamic (asterisks,
$m=0.5$) critical stresses $F_{cs}$ and $F_{cd}$ versus $A$. (b)
Theoretical (solid line, $m=0$) and numerical (squares, $m=0$;
asterisks, $m=0.5$) dislocation velocity vs.\ $F$ ($A=1$,
$N=25$).}
\label{fig3}
\end{center}
\end{figure}

To calculate $F_{cs}$, we extend the depinning calculations of
Ref. \onlinecite{carPRL01} to 2D systems. We redefine $u_{i,j}=
U_{i,j} + Fj$, insert $U_{i,j}= U_{i,j}(F,A) + v_{i,j}(t)$ in
Eq.\ (\ref{kkl}) with $m=0$ and expand the resulting equation
in powers of $v_{i,j}$, {\em about the stationary state}
$U_{i,j}(A,F)$ up to cubic terms. Subscripts in the resulting
equation can be numbered with a single one starting from the
point $i=j= -N$: $U_{i,j} = U_k$ and $v_{i,j} = v_k$, $k=i +
(j+N) (2N+1)$ for $i,j= -N,\ldots, N$. The resulting equation
can be written formally as $d{\bf v}/dt = {\cal M}(F){\bf
v} + {\cal B}({\bf v},{\bf v};F)$, where the vector ${\bf v}$
has components $v_k$. The linear stability of the stationary
state $U_k(A,F)$ depends on the eigenvalues of the matrix
${\cal M}(F)$. These eigenvalues are all real negative for $|F|
< F_{cs}$ whereas one of them vanishes at $|F|=F_{cs}$. This
criterion allows us to numerically determine $F_{cs}$ as a
function of $A$; see Fig.~\ref{fig3}(a). Notice that the
critical stress increases with $A$. Thus narrow core
dislocations ($A$ large) are harder to move. 

{\em Dislocation velocity}. Let us assume that $F>F_{cs}$ (the
case $F<-F_{cs}$ is similar). Then $v_{ij} = (F-F_c) j + \phi(t)
r_{ij}$ (plus terms that decay exponentially fast in time). The
procedure sketched in Refs.~\onlinecite{carPRL01,carPRE01} for
discrete 1D systems yields the amplitude equation $d\phi/dt
=\alpha +\beta\phi^2$. Here $\alpha= {\bf l}\cdot {\cal
N}(F_{cs})\, (F-F_{cs})/( {\bf l} \cdot {\bf r})$, ${\cal
N}_{ij} = A\, [\cos(U_{i,j+1}-U_{i,j} + F_c)- \cos(U_{i,j} -
U_{i,j-1} + F_c)]$ and $\beta = {\bf l}\cdot {\cal B}({\bf r},
{\bf r};F_{cs})/({\bf l}\cdot {\bf r})$, ${\bf l}$ and ${\bf
r}$ are the left and right eigenvectors of the matrix ${\cal
M}(F_{cs})$ corresponding to its zero eigenvalue (its  largest
one). From the amplitude equation, the approximate  dislocation
velocity is \cite{carPRL01}: $c\sim\sqrt{ \alpha\beta} /\pi = 
O(|F-F_{cs}|^{{1\over 2}})$. Numerically measured and
theoretically predicted dislocation velocities are compared in
Fig.~\ref{fig3}(b). Calculations in lattices of different sizes
yield similar results.

\begin{figure}
\begin{center}
\includegraphics[width=8cm]{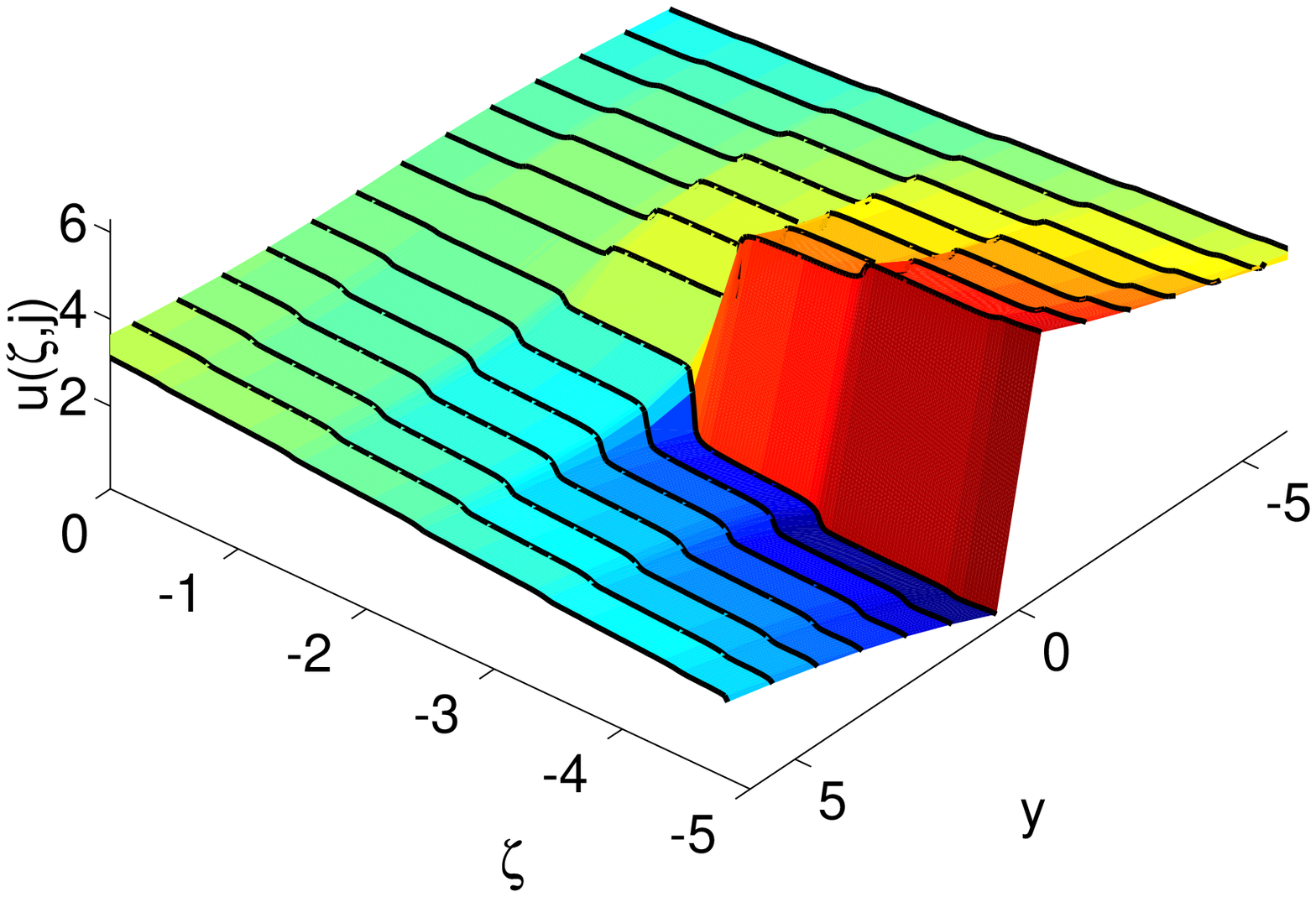}
\caption{Wave front profiles, $u_{i,j}(t)=u(\zeta,j)$,
$\zeta=i-ct$, $c>0$, near $F= F_{cs}$ for $A=3$, $m=0$ and
$N=25$.}
\label{fig4}
\end{center}
\end{figure}

How do we calculate numerically the dislocation velocity? This
is an important point for using the calculated dislocation
velocity as a function of stress in mesoscopic theories and a
few comments are in order. If we solve numerically Eq.\
(\ref{kkl}) with static NBC for $|F|> F_{cs}$, the velocity of
the dislocation increases as it moves towards the boundary. The
dislocation accelerates because we are using the far field of a
steady dislocation as boundary condition, instead of the (more
sensible) far field of a moving dislocation. However, the
latter is in principle unknown because we do not know the
dislocation speed. We will assume nevertheless that the
dislocation moves at constant speed $c$ once it starts moving,
as it would in a stressed infinite system. Then the correct
dislocation far field is $\theta(i- ct,j/\sqrt{A}) +Fj$. With
this far field in the NBC, Eq.\ (\ref{kkl}) has traveling wave
solutions $u_{i, j}(t)$ whose velocity can be calculated
self-consistently. How? By an iterative procedure that adopts
as initial trial velocity that of a dislocation subject to
static NBC as it starts moving. Near threshold, step-like
profiles are observed (see Fig.~\ref{fig4}), that become
smoother as $F$ increases. The profiles have been calculated by
following the trajectories of points with the same value of $i$
and different values of $j$ for $F> F_{cs}(A)$, according to
the formula $u_{i,j} (t) = u(\zeta, j)$, $\zeta=i-ct$. Notice
that the wave front profiles are kinks for $j< 0$ and antikinks
for $j\geq 0$. 

{\em Influence of fluctuations}. The original discrete model
contains both damping and fluctuation terms \cite{kkl}.
Fluctuation terms are appreciable only near $F_{cs}$, and
contribute an additive white noise term to the amplitude
equation. Due to this term, there is a small probability for
the dislocation to move even if $|F|< F_{cs}$ and $m=0$. The
resulting average velocity can be estimated by observing that
the potential of the corresponding Fokker-Planck equation is
cubic and it has a small barrier of height proportional to
$(F_{cs}-|F| )^{{3\over 2}}$. Then the exponentially small
velocity of the dislocation under the critical stress is the
reciprocal of the mean escape time from the barrier
\cite{vankampen}. Provided $(F_{cs}-|F|)\gg D$ (where $D$
measures the noise strength), we have $-\ln c\propto (F_{cs}
-|F|)^{{3\over 2}}/D$.

{\em Inertia and dynamic Peierls stress.} Inertia changes the
previous picture of dislocation motion in one important aspect:
the dislocations keep moving for an interval of stresses below
the static Peierls stress, $F_{cd}< |F|<F_{cs}$. On this stress
interval, stable solutions representing static and moving
dislocations coexist: to depin a static dislocation, we need 
$|F| > F_{cs}$. However if $|F|$ decreases below $F_{cs}$, a
moving dislocation keeps moving until $|F|<F_{cd}$; see Fig.
\ref{fig3}. Thus $F_{cd}$ represents the {\em dynamic Peierls
stress} of the dislocationÊ\cite{nab67}. Our theory therefore
yields the static and the dynamic Peierls stresses and the
velocity of a dislocation. 

{\em Interaction between edge dislocations}. Let us assume that
there are $N$ static edge dislocations at the points $(x_n,y_n)$
parallel to one dislocation at $(x_0,y_0)$, and that all
dislocations are separated from each other by distances of
order $L\gg 1$ (measured in units of the Burgers vector length,
$b$). We want to analyze whether the collective influence of the
$N$ distant dislocations can move that at $(x_0,y_0)$. This
problem is similar to that of deriving a reduced dynamics for
the centers of 2D vortices of Ginzburg-Landau equations subject
to their mutual influence \cite{neu90}. In the case of
dislocations, the existence of a pinning threshold implies that
the reduced dynamics is that of a single dislocation subject to
the mean field created by the others. We thus have a reduced
field dynamics, not particle dynamics as in the case of the
Ginzburg-Landau vortices.

Displacement vectors pose the problem of defining branch cuts
in the continuum (elastic) limit, which leads us to consider
instead the distortion tensor as a primary quantity
\cite{ll7}. For our discrete system, the distortion tensor has
nonzero components $w^{(1)}_{i,j} = u_{i+1,j}-u_{i,j}$ and
$w^{(2)} _{i,j} = \sin( u_{i,j+1}-u_{i,j})$ that become
$\epsilon\, (\partial u/\partial x)$ and $\epsilon\, (\partial
u/\partial y)$, respectively, in the continuum limit
$\epsilon=1/L\to 0$, $x-x_0=\epsilon i$, $y -y_0=\epsilon j$
finite. In the continuum limit, the distortion tensor of an
edge dislocation centered at the origin has nonzero components
$w^{(1)} = - \epsilon\sqrt{A}y/(A x^2+ y^2)$ and $w^{(2)} =
\epsilon\sqrt{A} x/(A x^2+y^2)$. If we have $N$ edge
dislocations at $(x_n,y_n)$, $1\leq n\leq N$, far from one at
$(x_0,y_0)$, the distortion tensor is sum of individual
contributions. Then the far field distortion tensor seen by the
dislocation at $(x_0,y_0)$ is:
\begin{eqnarray}
w^{(1)}_{i,j} = - {\sqrt{A}j\over Ai^2+j^2} + F_1 ,\quad \quad
w^{(2)}_{i,j} = {\sqrt{A}i\over Ai^2+j^2} + F_2 ,\label{c2}\\
F_1 =-\epsilon\sum_{n=1}^N {\sqrt{A} (y_0-y_n)\over A(x_0-x_n)^2
+(y_0-y_n)^2} + \ldots ,\label{c3}\\
F_2= \epsilon \sum_{n=1}^N {\sqrt{A} (x_0-x_n)\over A(x_0
-x_n)^2 +(y_0-y_n)^2} +\ldots .\label{c4}
\end{eqnarray}
The dislocation at $(x_0,y_0)$ moves if $F_2> F_{cs}(A)$. This
cannot be achieved as $\epsilon\to 0$ unless $N=O(1/\epsilon)$.
Then the sums in Eqs.\ (\ref{c3}) and (\ref{c4}) become
integrals. We define a static dislocation density $\rho(x,y)$
as the limit of $N^{-1} \sum_{n =1}^N \delta(x-x_n)
\delta(y-y_n)$ as $N\to\infty$. Then 
\begin{eqnarray}
F_1 =-\alpha \int_{-\infty}^\infty \int_{-\infty}^\infty
{\sqrt{A} (y_0-y) \rho(x,y)
\over A(x_0-x)^2 +(y_0-y)^2}\, dxdy ,\label{c5}\\
F_2= \alpha \int_{-\infty}^\infty\int_{-\infty}^\infty
{\sqrt{A}(x_0-x) \rho(x,y)
\over A(x_0 -x)^2 +(y_0-y)^2}\, dxdy, \label{c6}
\end{eqnarray}
where $\alpha=\epsilon N$ is the ratio of the total Burgers
vector to the mesoscopic length measuring average
interdislocation distance. As an example, let us assume that
$y_0=y_n=0$. Then $\rho= \rho(x)\delta(y)$ and $F_1=0$. Let us
assume that the dislocations are constrained by two obstacles at
$x_0=\pm l$ and subject to the same critical stress. Then the
critical dislocation density is $\rho(x)= [1-(\sqrt{A}F_{cs}/
\alpha)\, x]/[\pi \sqrt{l^2-x^2}]$, provided $\alpha>\sqrt{A}
F_{cs} l$ (cf.~\onlinecite{ll7}, page 127).

In conclusion, edge dislocations can be characterized as
traveling waves of discrete models. The dislocation far field
moves at a constant velocity equal to that of the dislocation
core. Static and dynamic Peierls stresses and the dislocation
velocity as a function of applied stress can be found
numerically (or analytically near critical stress in the
overdamped case) and adopted as the basis of a mesoscopic
theory \cite{gro00}. We have also shown that the interaction
between distant edge dislocations can be described in terms of a
continuous dislocation density. This field-theoretical reduced
description greatly  contrasts with the case of interacting
point vortices, which is completely described by the particle
dynamics of the vortex centers \cite{neu90}. Extension to fully
vectorial models and to other types of dislocation should follow
along similar lines.

We thank Joe Keller for helpful discussions. This work has been
supported by the MCyT grant BFM2002-04127-C02, by the  Third
Regional Research Program of the Autonomous Region of Madrid
(Strategic Groups Action), and by the European Union under grant
HPRN-CT-2002-00282.

%\newpage

%\clearpage

%\end{multicols}


\begin{thebibliography}{}
\bibitem{self}
The far field of a dislocation is not spatially homogeneous and
it should be rigidly and self-consistently displaced at the
velocity of the dislocation core.
\end{thebibliography}
\end{document}